\documentclass[12pt,twocolumn]{iopart}
\usepackage{graphicx}
\usepackage{cite}  
\begin{document}

\title{First-Principles Study on Electron Conduction in Sodium Nanowire}

\author{Yoshiyuki Egami$\dagger$, Takashi Sasaki$\dagger$, Tomoya Ono$\ddag$ and Kikuji Hirose$\dagger$}

\address{\dag\ Department of Precision Science and Technology, Osaka University, Suita, Osaka 565-0871, Japan.}

\address{\ddag\ Research Center for Ultra-Precision Science and Technology, Osaka University, Suita, Osaka 565-0871, Japan.}

\begin{abstract}
We present detailed first-principles calculations of the electron-conduction properties of a three-sodium-atom nanowire suspended between semi-infinite crystalline Na(001) electrodes during its elongation. Our investigations reveal that the conductance is $\sim 1~{\rm G_0}$ before the nanowire breaks and only one channel with the characteristic of the $3s$ orbital of the center atom in the nanowire contributes to the electron conduction. Moreover, the channel fully opens around the Fermi level, and the behavior of the channel-current density is insensitive to the structural deformation of the nanowire. These results verify that the conductance trace as a function of the electrode spacing exhibits a flat plateau at $\sim 1~{\rm G_0}$ during elongation.
\end{abstract}



\maketitle

\section{Introduction}
Metallic nanostructures have received great attention in the last decade. In particular, investigations of the electron-conduction properties of nanocontacts and nanowires are quite important since electronic devices are becoming highly integrated and miniaturized as nanoelectronics and nanotechnology progress. The size and structure dependences of the conduction properties are presently subjects of intensive research, and so far, numerous experimental studies on these dependences have been carried out\cite{krans2,krans1,hansen,costa2,costa1,ohnishi,ludoph,smit}. In the experiments, atomic-size nanowires are fabricated using scanning tunneling microscopy or mechanically controllable break junction techniques\cite{datta-ruiten}. A quantized conductance in units of ${\rm G_0}=2e^2/h$ is observed in the nanowires consisting of monovalent atoms, such as Na\cite{krans2,krans1}, Cu\cite{krans2,hansen,costa1}, Ag\cite{hansen,costa2} and Au\cite{hansen,costa1,ohnishi,ludoph,smit}, where $e$ is the electron charge and $h$ is Planck's constant.

On the theoretical side, there have been many first-principles calculations performed for nanowires sandwiched between electrodes\cite{nakamura,sim,lang1,lang2,kobayashi1,kobayashi2,havu,tsukamoto,brandbyge,palacios,lang3,wan,kobayashi3,furuya}, and it has been reported that the conductances of the Au or Na single-row nanowires are quantized in units of ${\rm G_0}$. However, since in some of those studies clusters\cite{nakamura,sim} or structureless jelliums\cite{lang1,lang2,kobayashi1,kobayashi2,havu,tsukamoto} have been substituted for the electrodes, unphysical effects from the electrodes are always of concern. For example, in the case of jellium electrodes, the conduction properties vary with the configuration of the interface between the nanowire and electrodes\cite{lang3,wan,kobayashi3,furuya}. In addition, although the nonequilibrium Green's function approach with a Gaussian basis is one of the popular techniques for investigating the conduction properties of nanostructures\cite{brandbyge,palacios}, precise descriptions of electronic states employing this method are difficult. For a strict study, one should employ more realistic models including semi-infinite crystalline electrodes. Recently, the overbridging boundary-matching (OBM) method\cite{obm,obm2}, which is a novel first-principles treatment of electron-conduction properties through a nanostructure suspended between semi-infinite crystalline electrodes, was developed by Fujimoto and Hirose, and several applications of it have been carried out\cite{otani,ono2,sasaki,egami}.

In this paper, we present a first-principles study on the electron-conduction properties of a three-sodium-atom nanowire suspended between semi-infinite crystalline Na(001) electrodes using the OBM method. The conductance trace as a function of electrode spacing is observed to exhibit a flat plateau at $\sim 1~\rm{G_0}$ before the nanowire breaks. By decomposing the current density into eigenchannels\cite{kobayashi1}, we also find that only one channel is relevant to the electron conduction. This channel arises essentially from the contribution of the $3s$ orbital of the center atom. In addition, the flat conductance trace before breaking is interpreted on the basis of the results, indicating that the characteristic of the conduction channel is negligibly affected by the structural deformation of the nanowire and that its transmission remains fully open around the Fermi level.

This paper is organized as follows: we describe the computational procedure in Sec.~2, our results are presented and discussed in Sec.~3, and we summarize our studies in Sec.~4.

\section{Computational Procedure}
\begin{figure}[ht]
\begin{center}
\includegraphics{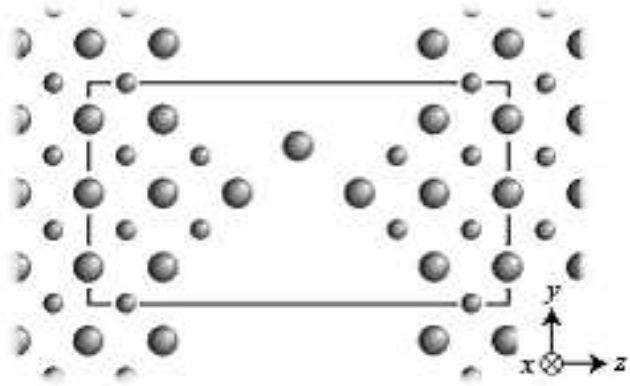}
\caption{Schematic view of calculation model. The three-sodium-atom nanowire is suspended between semi-infinite crystalline Na(001) electrodes. The rectangle represents the supercell employed for the structural optimization and the determination of the Kohn--Sham effective potential. The electrode spacing is elongated gradually.}
\label{fig1}
\end{center}
\end{figure}

Our first-principles calculation method is based on the real-space finite-difference approach\cite{cheliko1,cheliko2,ono} within the framework of the density functional theory\cite{hohen,kohn}. Figure~\ref{fig1} shows the schematic view of the calculation model. The nanowire is connected to the square bases made of sodium atoms and all of them are suspended between the semi-infinite crystalline Na(001) electrodes. The distance between the electrode surface and the basis, as well as that between the basis and the edge atom of the nanowire, is initially set to be $a_0/2$, where $a_0$ (=8.1~a.u.) is the lattice constant of sodium bulk; these are eventually determined by structural optimization.

The calculation procedure consists of two main steps. (i) The structural optimizations for the atomic geometry during the elongation are implemented using the conventional supercell of $3a_0 \times 3a_0 \times (2a_0+L_{es})$ represented by the rectangle in Fig.~\ref{fig1}, where $L_{es}$ is the electrode spacing. The supercell is imposed periodic boundary condition in $x$, $y$ and $z$ directions. The atoms in the nanowire and bases are relaxed by the structural optimization while the other atoms within the electrodes are fixed. The nine-point finite-difference formula is employed for the derivative arising from the kinetic-energy operator in the Kohn--Sham equation, and the grid spacing is taken to be 0.30~a.u. The ion cores are represented by the norm-conserving pseudopotential given by Troullier and Martins\cite{tmpp} and the exchange-correlation effects are treated by the local density approximation\cite{lda}.
(ii) The conductance is examined using the optimized atomic geometries obtained in step (i). Here, the central finite-difference formula, the local pseudopotential, the local density approximation and a grid spacing of 0.81~a.u. are employed. The Kohn--Sham effective potential is determined self-consistently using the conventional supercell shown by the rectangle in Fig.~\ref{fig1}.
In the calculations of the electron-conduction properties, the supercell is imposed the periodic boundary condition in $x$ and $y$ directions and the nonperiodic boundary condition in the $z$ direction. We have clarified that the enlargement of the supercell does not significantly change the electron conduction property. The global wave functions for infinitely extended states continuing from one electrode side to the other are determined by employing the OBM method. The electron transmission of the system at a zero bias limit is evaluated using the Landauer formula\cite{landauer}
\begin{equation}
\label{eqn1}
G(E_F)={\rm G_0}\sum_{i}T_i(E_F),
\end{equation}
where $G(E_F)$ is the conductance and $T_i(E_F)$ is the transmission of the $i$-th conduction channel at the Fermi level $E_F$.
The eigenchannels are investigated by diagonalizing the Hermitian matrix (${\bf T}^{\dagger}{\bf T}$)\cite{kobayashi1}, where {\bf T} is the transmission matrix.

\section{Results and Discussion}
\begin{figure*}[tb]
\begin{center}
\includegraphics{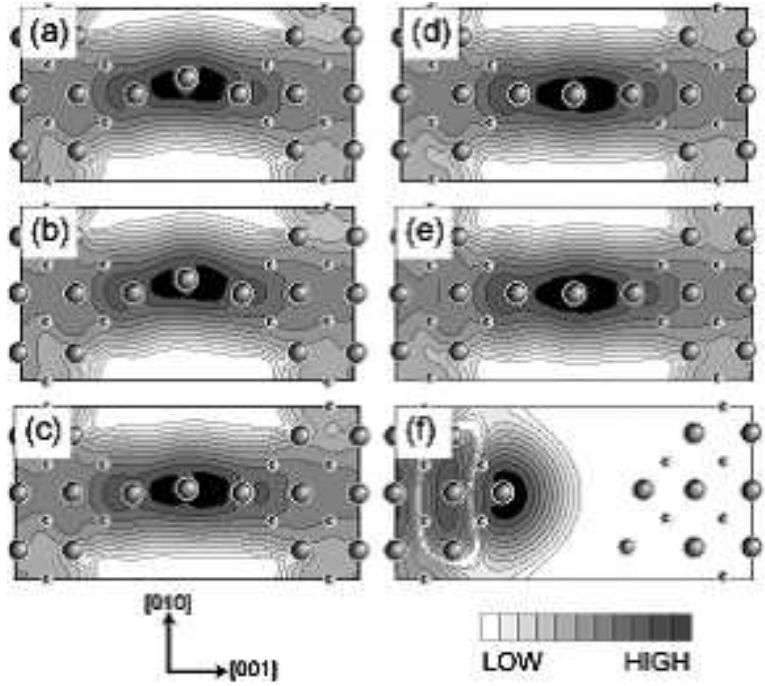}
\caption{Optimized geometries of nanowire and density distributions of electrons incident from left-hand electrode at the Fermi level. (a) $L_{es}=27.69$ a.u., (b) $L_{es}=28.37$ a.u., (c) $L_{es}=29.04$ a.u., (d) $L_{es}=29.72$ a.u., (e) $L_{es}=30.39$ a.u., and (f) $L_{es}=31.07$ a.u. Each contour represents twice or half the density of the adjacent contour curves. The lowest contour represents 9.93$\times 10^{-10}$ electron/$\mbox{bohr}^3$/eV.}
\label{fig2}
\end{center}
\end{figure*}

\begin{figure}[ht]
\begin{center}
\includegraphics{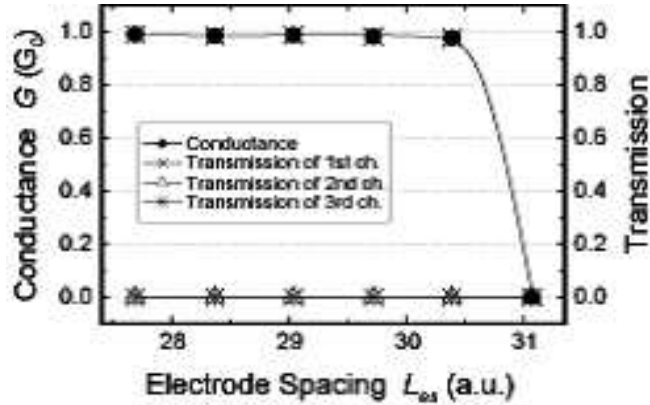}
\caption{Conductance and channel transmissions of three-sodium-atom nanowire as a function of electrode spacing.}
\label{fig3}
\end{center}
\end{figure}

The optimized geometries of the nanowire during stretching and the density distributions of the electrons incident from the left-hand electrode at the Fermi level are illustrated in Fig.~\ref{fig2}. The planes shown are perpendicular to the [100] direction and include the nanowire axis. The nanowire exhibits a bent structure below $L_{es}=29.04$~a.u. (Figs.~\ref{fig2}(a)~--~(c)). When elongated up to $L_{es}=29.72$~a.u., the nanowire manifests a geometrical transition from a bent structure to a straight one (Figs.~\ref{fig2}(d) and (e)), and finally the nanowire fractures at $L_{es}=31.07$~a.u. (Fig.~\ref{fig2}(f)).
We depict, in Fig.~\ref{fig3}, the conductance and the channel transmission as a function of the electrode spacing. The conductance is $\sim 1~\rm{G_0}$ and the value does not significantly change throughout the entire range of the electrode spacing below $L_{es}=30.39$~a.u. The conductance becomes zero after the nanowire distortion. These results of the conductance are in agreement with the results of the previous theoretical and experimental studies\cite{krans2,nakamura,kobayashi1}.
The details of the conduction channels can be described in terms of the decomposition of the incident electron-density distribution into eigenchannels. In every case, there is only one channel mainly contributing to the electron conduction. 

\begin{figure}[ht]
\begin{center}
\includegraphics{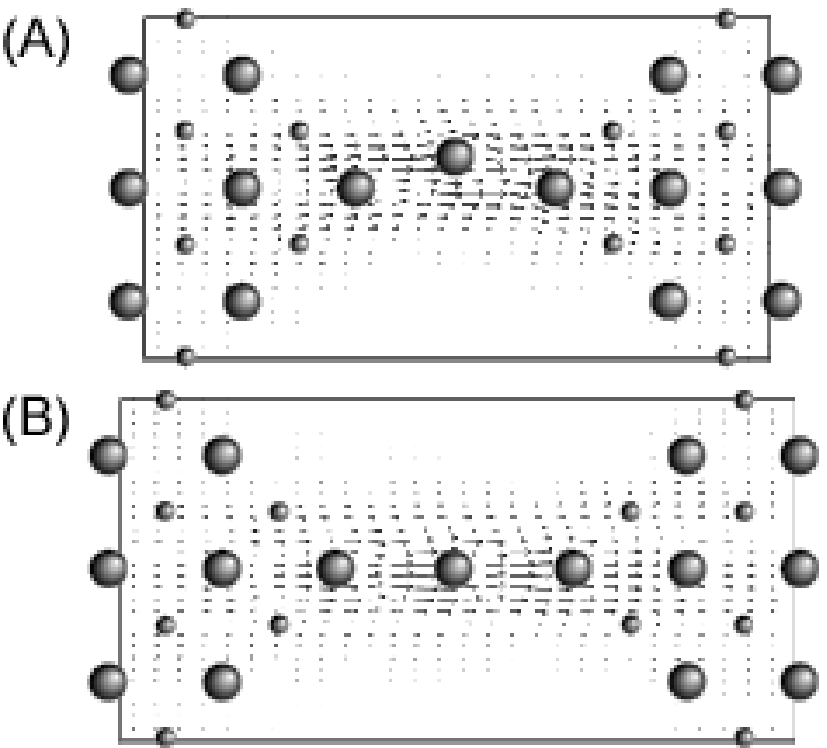}
\caption{Channel current-density distributions of electrons incident from left-hand electrode at the Fermi level for (A) bent nanowire ($L_{es}=28.37$~a.u.) and (B) straight nanowire ($L_{es}=30.39$~a.u.).}
\label{fig4}
\end{center}
\end{figure}

\begin{figure*}[ht]
\begin{center}
\includegraphics{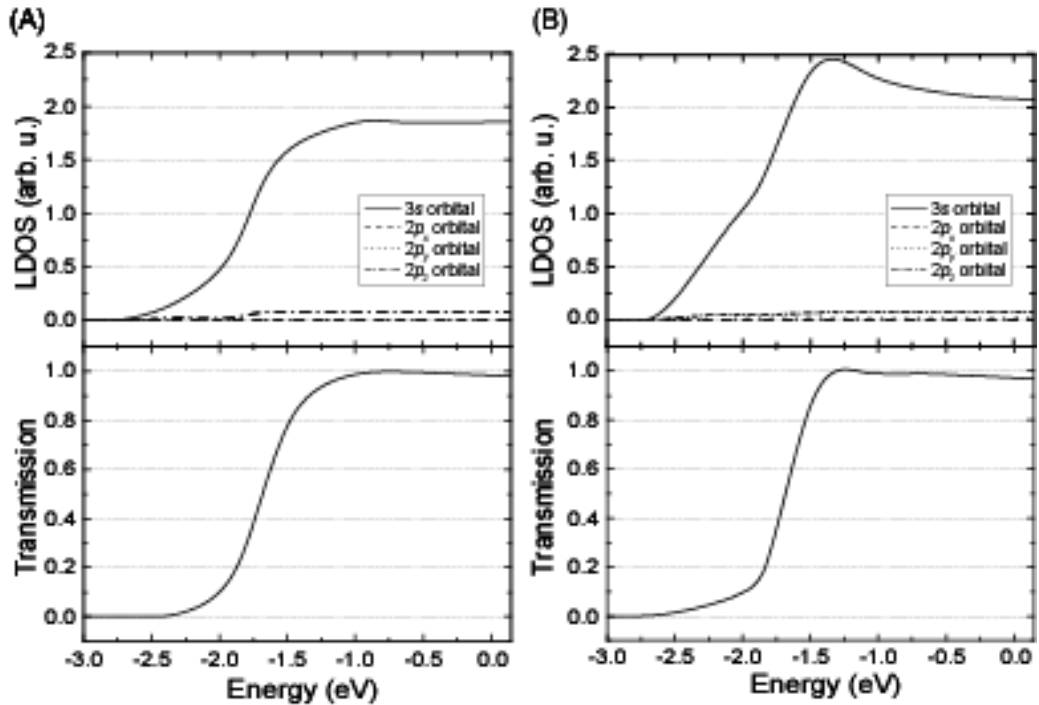}
\caption{LDOS around the center atom (upper panel) and channel transmission (lower panel) for the first channel as a function of the energy measured from the Fermi level. (A) Bent nanowire ($L_{es}=28.37$~a.u.) and (B) straight nanowire ($L_{es}=30.39$~a.u.).}
\label{fig5}
\end{center}
\end{figure*}

In Figs.~\ref{fig4}(A) and (B), the electronic current-density distributions for a bent nanowire ($L_{es}=28.37$~a.u.) and a straight one ($L_{es}=30.39$~a.u.), which correspond to configurations (a) and (e) in Fig.~\ref{fig2}, respectively, are shown. There is no marked difference between these current-density distributions; the behavior of the electronic current-density distributions is insensitive to the structural deformation of the nanowire before breaking.
In order to explore the characteristic of the current-density distributions in detail, we examine the local density of states (LDOS) of the conduction channels around the center atom of the nanowire and their transmissions. In Fig.~\ref{fig5}, the LDOS and the transmission of the first channel are plotted as a function of the energy measured from the Fermi level for (A) the bent nanowire ($L_{es}=28.37$~a.u.) and (B) the straight one ($L_{es}=30.39$~a.u.). In both cases, the $3s$ orbital of the center atom is dominant in electron conduction. This result supports the independence of current-density distributions from structural deformation, as shown in Fig.~\ref{fig4}. The peaks in the LDOS and the transmission are interpreted by the oscillation of the transmission probability as found in the one-dimensional penetration problem\cite{messiah}. The magnitude of the LDOS is mainly affected by the position of the center atom: in the case of the bent nanowire, since the center atom is not located on the central axis of the current flow, the magnitude of the LDOS becomes lower compared with the case of the straight nanowire. In addition, the LDOS and the channel transmission are fairly constant around the Fermi level. Because the conductance of the nanowire is determined by the characteristic of the conduction channel and its transmission at the Fermi level, our results that the channel current-density distribution is insensitive to the structural deformation of the nanowire and that its transmission is unity around the Fermi level are consistent with the experimental result that the conductance trace as a function of electrode spacing exhibits a flat plateau at $\sim 1~\rm{G_0}$ during elongation.

\section{Conclusion}
We performed the first-principles calculations of electron-conduction properties of a three-sodium-atom nanowire suspended between semi-infinite crystalline Na(001) electrodes. The conductance trace of the nanowire exhibits a flat plateau at $\sim 1~{\rm G_0}$ during stretching. For a more profound interpretation, we examined the current-density distributions of electrons incident from the left-hand electrode at the Fermi level and decomposed them into the atomic orbital of the center atom. There is no significant difference between the characteristics of the current flow of the bent nanowire and the straight one. Only one channel, which mainly has the character of the $3s$ orbital of the center atom, is almost fully open around the Fermi level. This result is consistent with the experimental conductance trace that has the flat plateau at $\sim 1~{\rm G_0}$ during elongation. In a future study, we will treat the nonlocal norm-conserving pseudopotential to more precisely investigate the conduction properties of the various nanowires.

\section{Acknowledgements} 
This research was supported by a Grant-in-Aid for the 21st Century COE ``Center for Atomistic Fabrication Technology'' and also by a Grant-in-Aid for Young Scientists (B) (Grant No. 14750022) from the Ministry of Education, Culture, Sports, Science and Technology. The numerical calculation was carried out with the computer facilities at the Institute for Solid State Physics at the University of Tokyo, and the Information Synergy Center at Tohoku University.

\end{document}